 \documentclass[final,3p,times]{elsarticle}

\usepackage{amssymb}
\usepackage{lipsum}

\journal{Machine Learning with Applications}

\begin{document}

\begin{frontmatter}



\title{The Butterfly Effect in Artificial Intelligence Systems:\\ Implications for AI Bias and Fairness}


\author[first]{Emilio Ferrara}
\affiliation[first]{organization={University of Southern California},
            addressline={Thomas Lord Department of Computer Science}, 
            city={Los Angeles},
            postcode={90007}, 
            state={CA},
            country={USA}}

\begin{abstract}
The Butterfly Effect, a concept originating from chaos theory, underscores how small changes can have significant and unpredictable impacts on complex systems. In the context of AI fairness and bias, the Butterfly Effect can stem from a variety of sources, such as small biases or skewed data inputs during algorithm development, saddle points in training, or distribution shifts in data between training and testing phases. These seemingly minor alterations can lead to unexpected and substantial unfair outcomes, disproportionately affecting underrepresented individuals or groups and perpetuating pre-existing inequalities. Moreover, the Butterfly Effect can amplify inherent biases within data or algorithms, exacerbate feedback loops, and create vulnerabilities for adversarial attacks. Given the intricate nature of AI systems and their societal implications, it is crucial to thoroughly examine any changes to algorithms or input data for potential unintended consequences. In this paper, we envision both algorithmic and empirical strategies to detect, quantify, and mitigate the Butterfly Effect in AI systems, emphasizing the importance of addressing these challenges to promote fairness and ensure responsible AI development.
\end{abstract}







\end{frontmatter}




\section{
Introduction
}
The Butterfly Effect, a fundamental concept in chaos theory, describes the sensitive dependence on initial conditions in nonlinear, dynamic systems. It was coined by American mathematician and meteorologist Edward Lorenz in the early 1960s (Lorenz, 1963). The Butterfly Effect suggests that small initial changes in complex, dynamic systems can result in significantly different and often unpredictable outcomes over time. The idea is best portraited by the popular saying that the flap of a butterfly's wings in Brazil could set off a chain of events leading to a tornado in Texas. While working on numerical weather prediction models, Lorenz observed that small changes in initial conditions led to drastically different long-term forecasts. The importance of the Butterfly Effect extends beyond meteorology and has found applications in various scientific disciplines, including physics, engineering, biology, and social sciences. In these disciplines, the Butterfly Effect is used to describe the consequences of small changes or perturbations in complex systems: this effect highlights the non-linear nature of complex systems as well as the importance of understanding their interconnectedness of various components and the challenges associated with accurate long-term predictions (Strogatz, 1994).

\subsection{Relevance of the Butterfly Effect to AI fairness and bias}
In the context of AI fairness and bias, the Butterfly Effect highlights the potential for small biases or skewed data inputs at various stages of algorithm development to result in significant and unexpected unfair outcomes. This phenomenon can manifest in various ways, such as small adjustments to input data, inherent biases within the data or algorithms themselves, shifts in data distributions, adversarial attacks, or feedback loops that amplify existing biases. The interconnected nature of AI systems, combined with their potential impact on society, makes it crucial to examine the Butterfly Effect's role in AI fairness and bias.

\subsection{Factors contributing to the Butterfly Effect in AI systems}
The relevance of the Butterfly Effect to AI fairness and bias lies in the observation that AI and machine learning (ML) systems are complex and interconnected, with multiple components contributing to their final decision-making process. These components include data, algorithms, and user interactions, among others. As a result, seemingly small changes or biases in these components can have a significant and potentially unforeseen impact on the fairness and bias of AI systems. Furthermore, seemingly small initial biases can propagate and cause large disparities. Vice versa, larger initial biases might possibly have less dramatic consequences in real applications. Several factors can contribute to the emergence of the Butterfly Effect on AI fairness and bias, summarized in Table~\ref{tab:factors}.

\begin{table}
\centering\small
\begin{tabular}{|p{2.75cm}|p{11cm}|p{1.5cm}|}
\hline
\textbf{Contributing Factor} & \textbf{Description} & \textbf{References} \\
\hline
High-dimensional input space & ML algorithms often operate on high-dimensional input data, which means they rely on a multitude of features to make decisions. Small perturbations in the input data, such as the removal or addition of features, can lead to vastly different model behavior and predictions. This sensitivity to input data can manifest as a Butterfly Effect, where slight adjustments result in significant and unintended consequences. & Barocas \& Selbst, 2016 \\
\hline
Nonlinearity and complexity of ML models & Many ML models, such as deep neural networks, are highly nonlinear and complex. This nonlinearity can make it challenging to predict how changes in input data or model parameters will affect the model's predictions. Consequently, biases or errors introduced during the training process may propagate and amplify, leading to biased and unfair outcomes. & Goodfellow et al., 2016 \\
\hline
Feedback loops and reinforcement of biases & ML systems can inadvertently create feedback loops that perpetuate and amplify biases. For example, if a biased ML system is used to generate new data, this new data may also be biased, reinforcing the system's existing biases over time. These feedback loops can create a Butterfly Effect, where small initial biases lead to increasingly biased outcomes as the system iterates. & Ensign et al., 2018 \\
\hline
Compounding effects of multiple components & AI systems often comprise multiple components, each potentially introducing its biases. These biases can interact and compound in unpredictable ways, leading to a Butterfly Effect where the overall system exhibits greater bias than any single component. & Friedler et al., 2019 \\
\hline
Local minima and distribution shifts & Saddle points in the loss landscape can stall optimization, causing algorithms to converge to suboptimal solutions that exacerbate fairness and bias issues. Distribution shifts, when test data differs from training data, can lead to poor model performance, unforeseen biases, and unfair outcomes. Both factors can trigger the Butterfly Effect on AI fairness and bias, amplifying minor discrepancies and yielding significant, unpredictable consequences. & Słowik \& Bottou, 2021; Jordan, 2023; Rezaei et al., 2021 \\
\hline
Adversarial attacks & By exploiting small perturbations or vulnerabilities in AI models, adversarial attacks can intentionally introduce subtle changes to input data or manipulate the model's decision boundaries, causing the model to produce significantly different and biased outcomes. Small alterations introduced by adversarial attacks can lead to substantial and unpredictable consequences in AI fairness and bias, thereby manifesting the Butterfly Effect in these systems. & Nanda et al., 2021 \\
\hline
\end{tabular}
\caption{Factors contributing to the Butterfly Effect in AI systems}
\label{tab:factors}
\end{table}

\section{
Examples of real-world emergence of the Butterfly Effect in AI systems
}
The Butterfly Effect can manifest in various ways in AI systems, leading to unintended consequences and exacerbating fairness and bias issues. Here, we discuss three real-world examples of the Butterfly Effect in AI systems.

\subsection{
Facial recognition technology
}
Facial recognition algorithms have become increasingly prevalent in various applications, from social media platforms to law enforcement. However, these algorithms can exhibit significant performance disparities across different demographic groups due to imbalanced training datasets. Small biases in the demographic representation of these datasets can lead to large differences in the algorithm's accuracy for different groups, resulting in biased outcomes that disproportionately affect underrepresented populations.

In a study by Buolamwini and Gebru (2018), they found that commercial facial recognition systems had higher error rates for darker-skinned and female subjects compared to lighter-skinned and male subjects. These disparities in performance can be attributed to the underrepresentation of specific demographic groups in the training data, leading to a Butterfly Effect in the fairness and bias of facial recognition systems.

\subsection{
Healthcare algorithms
}
AI and ML models are increasingly being used to support decision-making in healthcare, such as identifying high-risk patients and guiding treatment decisions. However, biases in historical data and model assumptions can lead to biased predictions, disproportionately impacting minority populations. 

Obermeyer et al. (2019) found that a widely-used commercial algorithm for predicting healthcare needs exhibited significant racial bias. The algorithm assigned lower risk scores to Black patients than White patients with similar health conditions, leading to disparities in access to care management programs. The study revealed that the algorithm relied on healthcare costs as a proxy for health needs, which inadvertently introduced bias due to racial differences in healthcare utilization. The small initial bias in the model's assumptions led to a Butterfly Effect, resulting in large disparities in the allocation of healthcare resources.

\subsection{
Hiring algorithms
}
AI-based recruiting tools are increasingly being employed to streamline the hiring process and identify qualified candidates. However, these tools can perpetuate and amplify existing biases in the hiring process, leading to unfair outcomes and exacerbating societal inequalities (Raghavan et al. 2020).

In 2018, it was reported that an AI recruiting tool showed gender bias, favoring male candidates over female candidates for technical roles (Dastin, 2018). The bias emerged due to the training data, which consisted primarily of resumes submitted to the company over a ten-year period and reflected a male-dominated applicant pool. Additionally, the algorithm might have penalized resumes containing words associated with women, such as ``women's'' in phrases like ``women's chess club.'' The biases present in the training data and the use of gender-biased proxies led to a Butterfly Effect, resulting in a biased AI system that perpetuated gender inequality in hiring.

\subsection{
Large Language Models
}
Large language models, such as GPT-4 or Bard, could be significantly impacted by the Butterfly Effect. The training process for these models involves learning from vast amounts of text data, making them susceptible to minor changes in input data or algorithmic processes. A seemingly insignificant alteration in the training data, such as a slightly skewed representation of a particular demographic or viewpoint, can lead to substantial and unexpected biases in the model's output. Additionally, adversarial attacks, distributional shifts between training and test data could also lead to unforeseen outcomes. Hence, the Butterfly Effect could manifest in large language models, resulting in outputs that propagate and amplify pre-existing biases or inaccuracies -- see Weidinger et al. (2021); Ferrara (2023).

\section{
Manifestations of the Butterfly Effect in AI systems
}
\begin{figure}[h]
    \centering
    \includegraphics[width=\columnwidth, clip, trim=0 35 0 0]{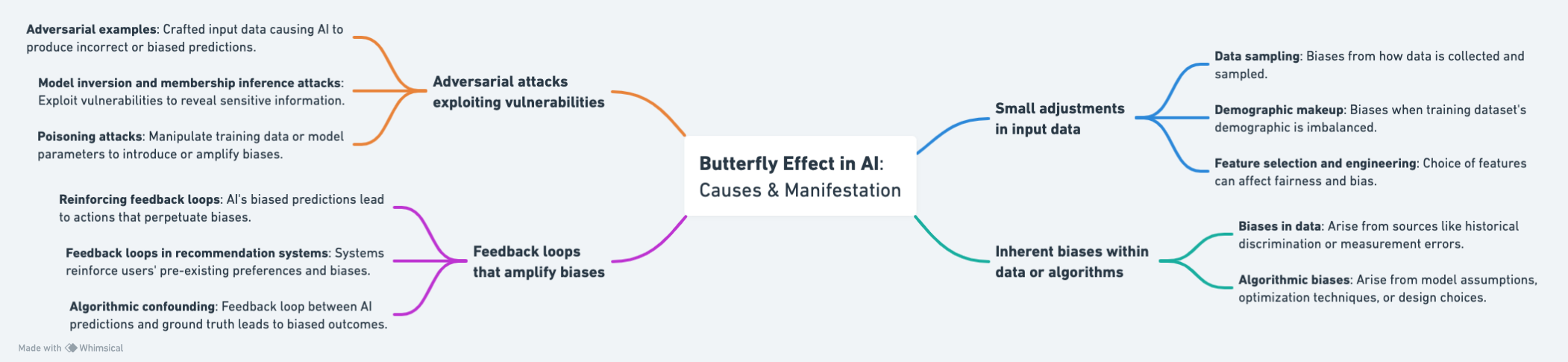}
    \caption{Root causes and associated manifestations of the Butterfly Effect in AI systems}
    \label{fig:3}
\end{figure}

We explore the various manifestations of the Butterfly Effect on AI fairness and bias. These manifestations include small adjustments in input data, inherent biases within data or algorithms, feedback loops that amplify biases, and adversarial attacks exploiting vulnerabilities. By understanding these different ways in which the Butterfly Effect can impact AI systems, we can better identify potential sources of unfairness and bias and develop appropriate strategies to mitigate their effects, ensuring that AI systems produce fair and unbiased outcomes. Figure~\ref{fig:3} summarizes the causes and manifestations of the Butterfly Effect in AI systems presented in detail next.

\subsection{Small adjustments in input data}
Small adjustments in input data can significantly impact the fairness and bias of AI systems, as these systems rely on vast amounts of data to make decisions. These adjustments can manifest in various ways, including changes in data sampling, demographic makeup, and feature selection, among others. This sensitivity to input data can lead to the Butterfly Effect, where minor alterations in input data result in significant and unintended consequences in the fairness and bias of AI systems.

\subsubsection{
Data sampling
}
The manner in which data is collected and sampled can introduce biases in AI systems. If the data collection process is not carefully designed, it can lead to underrepresentation or overrepresentation of certain groups, which can in turn affect the fairness and bias of the AI system (Shankar et al., 2020). For example, non-representative sampling from a population can result in skewed datasets, causing AI systems to perform poorly for underrepresented groups.

\subsubsection{
Demographic makeup
}
AI systems may exhibit biases when demographic makeup within the training dataset is imbalanced. For instance, if a particular demographic group is underrepresented in the training data, the AI system may not generalize well to that group, leading to biased and unfair outcomes (Buolamwini \& Gebru, 2018). In facial recognition technology, the underrepresentation of certain demographic groups in training datasets can lead to disparities in the algorithm's accuracy for those groups. Ensuring that training datasets are representative of diverse populations is crucial to mitigate the Butterfly Effect and ensure the fairness of AI systems.

\subsubsection{
Feature selection and engineering
}
Feature selection and engineering play a crucial role in shaping the behavior of AI systems. The choice of features used as inputs can significantly affect the fairness and bias of AI models (Chouldechova, 2017). For example, using features that are proxies for protected attributes, such as race or gender, can introduce bias into AI systems, even if the protected attributes themselves are not explicitly used. Additionally, the omission of important features that capture relevant information about the population may result in biased models that do not adequately account for differences between groups.

In conclusion, small adjustments in input data, such as data sampling, demographic makeup, and feature selection, can have a profound impact on the fairness and bias of AI systems. To mitigate the Butterfly Effect and ensure that AI systems promote fairness and equity, it is essential to carefully curate and preprocess input data to minimize biases and accurately represent diverse populations.

\subsection{
Inherent biases within data or algorithms
}
We delve into the inherent biases within data or algorithms that can cause the Butterfly effect on AI fairness and bias. These biases can originate from various sources, such as biased data collection processes or unintentional biases embedded in algorithm design.

Inherent biases within data or algorithms can lead to the Butterfly Effect on AI fairness and bias. Data biases can arise from historical discrimination, measurement errors, or other systemic issues affecting the data-generating process, while algorithmic biases can emerge from model assumptions, optimization techniques, or other design choices. These biases can propagate and compound throughout the system, leading to significant and unintended consequences.

\subsubsection{
Biases in data
}
Data biases can emerge from various sources, such as historical discrimination or measurement errors. For instance, historical biases present in training data can lead AI systems to perpetuate or exacerbate existing inequalities. An example of this phenomenon is seen in the COMPAS recidivism risk assessment tool, which was found to have disparate impacts on different racial groups due to biases present in the training data (Angwin et al., 2016). Another source of data bias is measurement error, which occurs when variables in the dataset do not accurately represent the underlying constructs they are meant to capture. Measurement errors can introduce biases in AI systems, leading to unfair decision-making (Dressel \& Farid, 2018).

\subsubsection{
Algorithmic biases
}
Algorithmic biases can arise from various aspects of the AI system, such as model assumptions, optimization techniques, or other design choices. Model assumptions, like the choice of a linear model or the assumption of independence between features, can introduce biases if they do not accurately reflect the underlying data-generating process (Berk et al., 2018). Optimization techniques, such as regularization or the choice of a loss function, can also lead to biases if they prioritize certain objectives over others (Kearns et al., 2018). Lastly, other design choices, like the selection of hyperparameters or the choice of an ensemble method, can introduce biases into AI systems, which can propagate and compound over time (Grgić-Hlača et al., 2018).

\subsection{Feedback loops that amplify biases}
Feedback loops can amplify biases in AI systems, leading to the Butterfly Effect compromising fairness and bias. Feedback loops occur when the output of an AI system influences its future inputs, reinforcing and magnifying biases over time. This can result in a self-perpetuating cycle of unfair outcomes that disproportionately impact certain groups.

\subsubsection{Reinforcing feedback loops}
Reinforcing feedback loops can occur when an AI system's biased predictions lead to actions that further perpetuate the initial biases. For example, predictive policing algorithms that rely on historical crime data can create a feedback loop by directing law enforcement resources to areas with higher reported crime rates (Lum \& Isaac, 2016). If certain groups or neighborhoods are disproportionately targeted due to historical biases in the data, the increased police presence can lead to more arrests and crime reports, which in turn reinforce the initial biases in the AI system.

\subsubsection{Feedback loops in recommendation systems}
Recommendation systems are another example where feedback loops can amplify biases. These systems often rely on user data to provide personalized recommendations, which can create a filter bubble that reinforces users' pre-existing preferences and biases (Nguyen et al., 2014). In turn, this can lead to biased content exposure and a lack of diversity in the information that users are exposed to, perpetuating existing social biases and contributing to polarization (Pariser, 2011).

\subsubsection{Algorithmic confounding}
Algorithmic confounding is a phenomenon where a feedback loop between an AI system's predictions and the ground truth it aims to predict leads to biased and unfair outcomes. This can occur when the AI system's biased predictions influence the data used to evaluate its performance, making it difficult to disentangle the true effect of the AI system from the biases present in the data. In such cases, biased algorithms may appear to perform well due to the confounding effect, reinforcing the initial biases and leading to a self-perpetuating cycle of unfair outcomes.

To address feedback loops and mitigate their amplifying effect on biases, it is essential to carefully consider the potential consequences of AI systems' predictions on their future inputs and design mechanisms for monitoring and correcting biases as they emerge over time.

\subsection{Adversarial attacks exploiting vulnerabilities}
Adversarial attacks can exploit vulnerabilities in AI systems, leading to the Butterfly effect that compromises fairness and bias. These attacks involve the intentional manipulation of input data or model parameters to cause an AI system to produce biased, unfair, or otherwise undesirable outcomes. By exploiting the sensitivity of AI systems to small changes, adversarial attacks can have significant and unpredictable consequences on fairness and bias.

\subsubsection{Adversarial examples}
Adversarial examples are carefully crafted input data designed to cause an AI system to produce incorrect or biased predictions (Szegedy et al., 2013). These examples can be generated by adding small, imperceptible perturbations to the input data, which can result in vastly different predictions due to the Butterfly Effect. Adversarial examples can be particularly problematic for fairness and bias, as they can be used to target specific demographic groups or individuals, leading to discriminatory outcomes (Sharif et al., 2016).

\subsubsection{Model inversion and membership inference attacks}
Model inversion and membership inference attacks can exploit the vulnerabilities of AI systems to reveal sensitive information about the training data or individuals within the dataset (Fredrikson et al., 2015; Shokri et al., 2017). These attacks can lead to the Butterfly Effect compromising fairness and bias by revealing disparities in the demographic makeup of the training data or exposing biases in the AI system's predictions. Furthermore, the knowledge gained from these attacks can be used to craft more sophisticated adversarial examples, further exacerbating the impact of the Butterfly Effect on fairness and bias.

\subsubsection{Poisoning attacks}
Poisoning attacks involve the manipulation of the training data or model parameters to introduce or amplify biases in an AI system (Biggio et al., 2012). By injecting carefully crafted examples into the training data or modifying the model parameters, adversaries can exploit the Butterfly Effect to cause an AI system to produce biased, unfair, or otherwise undesirable outcomes. Poisoning attacks can be particularly challenging to detect and mitigate, as the perturbations introduced by the attacker can be subtle and difficult to distinguish from natural variations in the data.

To defend against adversarial attacks and mitigate their impact on fairness and bias, it is essential to develop robust AI systems that can withstand small perturbations in the input data or model parameters (Madry et al., 2017). Additionally, regular monitoring and evaluation of AI systems for fairness and bias, as well as the implementation of privacy-preserving techniques, can help prevent adversaries from exploiting the Butterfly Effect to compromise the fairness and integrity of AI systems.

\subsection{Strategies to Mitigate the Butterfly Effect on AI fairness and bias}
\begin{figure}[h]
    \centering
    \includegraphics[width=\columnwidth, clip, trim=0 50 0 0]{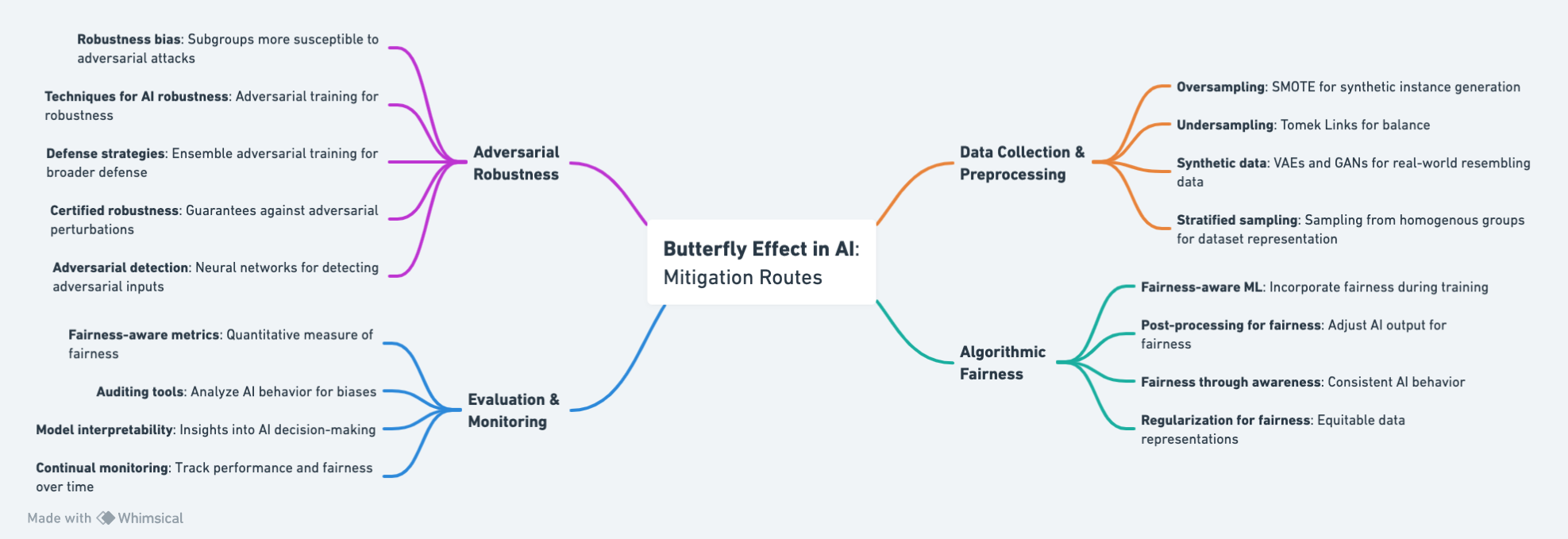}
    \caption{Strategies to Mitigate the Butterfly Effect on AI fairness and bias}
    \label{fig:4}
\end{figure}

Next, we discuss various strategies to mitigate the Butterfly Effect on AI fairness and bias. These strategies encompass diverse aspects of AI system development, ranging from data collection and preprocessing to algorithmic fairness, evaluation and monitoring, and adversarial robustness. By employing these strategies, researchers and practitioners can work towards addressing the potential unintended consequences arising from small changes in input data or algorithmic design and ensure that AI systems are more transparent, accountable, and fair. Figure~\ref{fig:4} summarizes the mitigation strategies presented in detail next.

\subsubsection{Data Collection and Preprocessing}
Creating balanced and representative datasets is crucial for mitigating the Butterfly Effect on AI fairness and bias. Several techniques can be employed to ensure that datasets are balanced and accurately represent the population of interest.

\begin{enumerate}
\item 
\textbf{Oversampling minority classes}: Oversampling involves creating copies of instances from minority classes to balance the class distribution. One well-known technique is the Synthetic Minority Over-sampling Technique (SMOTE), which generates synthetic instances of minority classes by interpolating between existing instances (Chawla et al., 2002). SMOTE can help alleviate the problem of overfitting associated with simple oversampling and lead to better performance in terms of fairness and generalization.
\item 
\textbf{Undersampling majority classes}: Undersampling involves removing instances from majority classes to balance the class distribution. One effective undersampling technique is Tomek Links, which removes majority class instances that are close to the decision boundary (Kubat \& Matwin, 1997). By removing these instances, the decision boundary becomes less sensitive to small changes in the data, mitigating the Butterfly Effect.
\item 
\textbf{Synthetic data generation}: Synthetic data generation can be used to create new, artificial instances for underrepresented classes, ensuring that the dataset is representative of the population of interest. Techniques such as Variational Autoencoders (VAEs) and Generative Adversarial Networks (GANs) can generate high-quality synthetic data that closely resembles the real-world distribution of the data, reducing the impact of small changes in the data on AI fairness and bias (Frid-Adar et al., 2018).
\item 
\textbf{Stratified sampling}: Stratified sampling is a method of sampling that involves dividing the population into homogenous groups, or strata, and sampling instances from each stratum in proportion to the size of the stratum. This technique can help ensure that the dataset is representative of the population of interest and reduce the sensitivity of AI systems to small changes in the data.
\end{enumerate}

\subsubsection{Algorithmic Fairness}
Algorithmic fairness is a critical aspect of machine learning that focuses on ensuring equitable treatment across different groups, mitigating biases, and reducing the sensitivity of models to minor changes in input data.

\begin{enumerate}
\item 
\textbf{Fairness-aware machine learning}: Fairness-aware machine learning aims to incorporate fairness constraints during the training process to minimize disparate treatment and disparate impact on different groups. Zafar et al. (2017) propose a convex optimization formulation for learning a classifier that satisfies various notions of fairness while maintaining high accuracy. The method minimizes the difference in correlations between the classifier's predictions and the sensitive attribute (e.g., race, gender) across different groups. By incorporating fairness constraints during the training process, this approach helps to mitigate the Butterfly Effect by reducing the sensitivity of the model to small changes in the input data.

\item 
\textbf{Post-processing methods for achieving fairness}: Post-processing techniques focus on adjusting the output of a trained model to ensure fairness. Hardt et al. (2016) propose a method that learns a transformation of the classifier's predictions to satisfy the equalized odds criterion (equal true positive and false positive rates across groups). The method requires no retraining of the original classifier and guarantees the best achievable trade-off between accuracy and fairness. By adjusting the model's output, post-processing methods can help to mitigate the Butterfly Effect by compensating for biases that may have been introduced during the training process.

\item 
\textbf{Fairness through awareness}: Dwork et al. (2012) introduce the concept of fairness through awareness, which requires that any two individuals who are similar with respect to a specific task should be treated similarly by the algorithm. They propose a Lipschitz condition on the classifier's behavior, ensuring that the model is less sensitive to small changes in the input data. Fairness through awareness can help to mitigate the Butterfly Effect by constraining the model's behavior to be consistent and less influenced by small perturbations in the data.

\item 
\textbf{Regularization for fairness}: Regularization techniques can be used to encourage fairness in AI models. Zhao et al. (2019) propose a method that incorporates a fairness-aware regularization term in the objective function during training. This term penalizes the difference between the predicted outcomes for different groups, encouraging the model to learn more equitable representations of the data. Regularization for fairness can help to mitigate the Butterfly Effect by discouraging the model from relying on small differences in the input data that may lead to unfair outcomes.
\end{enumerate}

\subsubsection{Evaluation and Monitoring}
We emphasize the importance of assessing, understanding, and continuously overseeing the fairness and behavior of AI systems to detect and address potential biases and unintended consequences due to the Butterfly Effect.

\begin{enumerate}
\item 
\textbf{Fairness-aware performance metrics}: Evaluating the fairness of AI models requires metrics that capture the disparate impact on different groups. Verma \& Rubin (2018) discuss a set of fairness-aware performance metrics, such as demographic parity, equalized odds, and equal opportunity. These metrics provide a quantitative measure of the difference in outcomes for different groups, helping to identify potential bias and unfairness in AI models. By using fairness-aware performance metrics, practitioners can monitor the impact of small changes in the data or model and identify instances where the Butterfly Effect leads to unintended consequences.

\item 
\textbf{Auditing tools for AI systems}: Auditing tools can help to systematically analyze the behavior of AI systems to identify potential biases, fairness violations, and other issues. Grgić-Hlača et al. (2018) present a method for auditing black-box models that involves perturbing the input data to explore the model's sensitivity to different features, specifically focusing on protected attributes. By systematically analyzing the model's behavior, auditing tools can help to identify potential Butterfly Effects and inform subsequent mitigation strategies.

\item 
\textbf{Model interpretability for fairness}: Model interpretability techniques can provide insights into the decision-making process of AI models, allowing for better scrutiny of potential bias and unfairness. Ribeiro et al. (2016) propose Local Interpretable Model-agnostic Explanations (LIME), a method for explaining the predictions of any classifier by approximating it locally with an interpretable model. By understanding the underlying factors that influence a model's decisions, practitioners can detect potential Butterfly Effects and address them through appropriate interventions.

\item 
\textbf{Continual monitoring and feedback}: Continual monitoring and feedback involve tracking the performance and fairness of AI systems over time, as well as collecting feedback from users to identify potential issues. Mitchell et al. (2018) discuss the importance of actively soliciting user feedback to identify biases, unfairness, and other issues that may not be apparent from traditional evaluation metrics. Continual monitoring and feedback can help to uncover Butterfly Effects that emerge over time or as a result of changing data distributions and user contexts.
\end{enumerate}

\subsubsection{Adversarial Robustness}
We delve into the challenges and solutions associated with ensuring AI models remain fair and unbiased when faced with adversarial attacks.

\begin{enumerate}
\item 
\textbf{Robustness bias}: Nanda et al. (2021) suggested that relying solely on traditional notions of fairness based on a model's outputs may not be enough when models are susceptible to adversarial attacks. To measure robustness bias, they proposed two methods and performed an empirical investigation on state-of-the-art deep neural networks using commonly employed real-world fairness datasets. Their findings demonstrated that subgroups, which are categorized by sensitive attributes like race and gender, are less robust and therefore more susceptible to adversarial attacks. It is important to consider robustness bias when evaluating real-world systems that rely on deep neural networks to make decisions (Nanda et al., 2021).

\item 
\textbf{Techniques for improving AI model robustness}: To mitigate the Butterfly Effect on AI fairness and bias caused by adversarial attacks, it is essential to develop models that are robust against such perturbations. Madry et al. (2017) propose a training framework based on adversarial training, which aims to minimize the worst-case loss over a set of allowed perturbations. By training the model to be robust against adversarial examples, this approach helps to improve the model's robustness against small changes in the input data, reducing the risk of unintended consequences due to the Butterfly Effect.

\item 
\textbf{Defense strategies against adversarial attacks}: Defending against adversarial attacks is crucial for ensuring the robustness of AI models and mitigating the Butterfly Effect. Tramèr et al. (2017) propose a defense strategy called ensemble adversarial training, which augments the training data with adversarial examples generated by an ensemble of models. This approach helps to increase the diversity of the adversarial examples used during training, making the model more robust against a broader range of attacks. By defending against adversarial attacks, this approach helps to reduce the impact of the Butterfly Effect on AI fairness and bias.

\item 
\textbf{Certified robustness}: Certified robustness aims to provide guarantees on the model's behavior under adversarial perturbations, ensuring that small changes in the input data do not lead to significant changes in the model's output. Cohen et al. (2019) introduce randomized smoothing, a technique that provides provable robustness guarantees for classifiers against l2-norm bounded adversarial perturbations. By providing certified robustness, this approach helps to mitigate the Butterfly Effect by ensuring that the model's behavior is stable under small perturbations in the input data.

\item 
\textbf{Adversarial detection}: Detecting adversarial examples is an essential step in defending against adversarial attacks and mitigating the Butterfly Effect on AI fairness and bias. Metzen et al. (2017) propose a method for detecting adversarial examples by training a separate neural network to distinguish between clean and adversarial inputs. By detecting adversarial examples, this approach helps to prevent unintended consequences due to the Butterfly Effect, ensuring that the model's behavior remains fair and unbiased.
\end{enumerate}

\section{Conclusions}

Throughout this paper, we have explored the Butterfly Effect's role in AI fairness and bias, a phenomenon rooted in chaos theory where small changes can lead to significant and unpredictable effects on complex systems. The manifestations of the Butterfly Effect in AI systems can arise from small adjustments in input data, inherent biases within data or algorithms, feedback loops that amplify biases, and adversarial attacks exploiting vulnerabilities.

Given the pervasive nature of AI systems and their increasing impact on various aspects of society, understanding the Butterfly Effect is crucial for ensuring fairness and minimizing unintended consequences. We have outlined a set of mitigation strategies that encompass data collection and preprocessing, algorithmic fairness, adversarial robustness, and continuous evaluation and monitoring. 

Implementing these strategies can help researchers and practitioners develop AI systems that are more transparent, accountable, and fair, ultimately promoting equitable outcomes and fostering trust in these systems. By rigorously scrutinizing the potential Butterfly Effect's role in AI systems and proactively working to mitigate the negative consequences, we can better ensure that AI technologies serve the greater good and contribute positively to societal progress.













\end{document}